\documentclass[11pt,a4paper]{article}
\usepackage[letterpaper, portrait, margin=1.15in]{geometry}
\usepackage{setspace}
\usepackage{amsmath}
\usepackage{verbatim}
\usepackage{float}
\usepackage{graphicx}
\newcommand{\cmmnt}[1]{}
\usepackage{algorithm,eqparbox}
\usepackage[noend]{algpseudocode}
\newcommand{\commentsymbol}{//}

\algrenewcommand\algorithmiccomment[1]{\hfill \commentsymbol{} #1}

\algrenewcommand\algorithmicindent{1.0em}
\makeatletter
\algnewcommand{\LineComment}[1]{\Statex \hskip\ALG@thistlm \(\commentsymbol{}\) #1}
\makeatother
\def\BState{\State\hskip-\ALG@thistlm}
\makeatother

\title{Encoding and Selecting Coarse-Grain Mapping Operators with Hierarchical Graphs}
\author{Maghesree Chakraborty, Chenliang Xu, Andrew D. White}

\begin{document}
\maketitle

\begin{abstract}
Coarse grain (CG) molecular dynamics (MD) can simulate systems inaccessible to fine grain (FG) MD simulations. A CG simulation decreases the degrees of freedom by mapping atoms from an FG representation into agglomerate CG particles. The FG to CG mapping is not unique. Research into systematic selection of these mappings is challenging due to their combinatorial growth with respect to the number of atoms in a molecule. Here we present a method of reducing the total count of mappings by imposing molecular topology and symmetry constraints. The count reduction is illustrated by considering all mappings for nearly 49,889 molecules. The resulting number of mapping operators is still large, so we introduce hierarchical graphs which encode multiple CG mapping operators. The encoding method is demonstrated for methanol and a 14-mer peptide. This encoding provides a foundation to perform automated mapping selection. \end{abstract}
\section{Introduction}
The length-scale of all-atom molecular dynamics (MD) simulation is fundamentally limited due to the large atom numbers of complex multiscale phenomena. One way to address this is through coarse-grain (CG) MD simulation, which maps a fine-grain (FG) or all-atom simulation to a lower resolution model with fewer degrees of freedom. CG simulation enables the study of larger length-scale systems, such as multi-protein complexes\cite{Morriss-Andrews2014,Kawaguchi2017,Ingolfsson2014}, lipid bilayers\cite{Arnarez2015,Morriss-Andrews2014,Ingolfsson2014}, nucleic acids\cite{Riniker2012a,Kmiecik2016,McCullagh2016,Ingolfsson2014,Becker2007} and polymers\cite{Milano2005,Huang2017,Aramoon2016}. The choice of mapping from FG to CG is not unique and thus continues to be an active area of research\cite{Riniker2012a}. Many CG mappings rely on chemical intuition\cite{Cao2015} or follow rules tailored for specific classes of molecules\cite{Milano2005,Izvekov2004,Vishnyakov2012}.

Mapping atoms into CG particles happens to be analogous to a method found in video processing called segmentation. Video segmentation involves merging similar voxels (volumetric pixels) into supervoxels\cite{Grundmann2010} to facilitate easier analysis of the video. Both CG and video segmentation involve dimensionality reduction to remove superfluous details while retaining salient features of the high resolution representation. This work is motivated by the successful application of graph-based methods in video segmentation\cite{Xu2013,Sharon2006,Xu2012a}. 

Specific work on CG mapping theory includes work from Zhang and Voth \cite{Zhang2008, Zhang2009, Zhang2010} who designed CG mapping operators of biomacromolecules such as proteins, ribosomes, and actin filaments by preserving the ``essential dynamics''\cite{Amadei1993} of the simulation. They made the optimization of the CG operator tractable by limiting analysis to one conformation of the FG system and through the use of a potential energy function that does not require iterative optimization (an elastic-network model).
 Cao and Voth worked on theory for the centering functions, specifically comparing mass centering and center of charge\cite{Cao2015}. They found that center of charge gave better preservation of inter-molecular structure in four example systems. Wagner et al.\cite{Wagner2016} developed theory that can generalize centering functions to arbitrary thermodynamic quantities. Marrink et al.\cite{Marrink2007} developed a hand-tuned mapping operator based on grouping four heavy atoms. Although this approach was empirical, their work showed that consistently using the same number of heavy atoms when grouping (fixed-size segmentation) is a valid simplification. 

Here we introduce the concept of encoding multiple CG mappings into a hierarchical graph. The hierarchical graph enables automated selection of CG mappings including multi-resolution mappings\cite{Brini2013,Bond2007}. We will first introduce a theory of CG mapping operators including enumeration formulas for CG mapping operators and a method of constructing the hierarchical graph. The theory is followed by examples to demonstrate these techniques on model systems. The capability of the hierarchical graph to encode CG mappings is tested and specific mappings are evaluated based on their ability to preserve radial distribution and velocity autocorrelation functions. 

\section{Theory}
A CG model is specified by a mapping operator, \textbf{M}, and a CG-potential U($\vec{R}$). \textbf{M} establishes the relationship between FG coordinate space, $\vec{r}$, and CG coordinate space, $\vec{R}$. Here, \textbf{M} is represented as a matrix of dimension $N_{\vec{R}}\times N_{\vec{r}}$ where $N_{\vec{r}}$ and $N_{\vec{R}}$ are number of FG particles and number of CG particles respectively. The coordinates of the $i^{\text{th}}$ CG particle can be written as $\vec{R_i}=\sum_{j=1}^{N_{\vec{r}}} M_{ij}\vec{r_j}$. $M_{ij}$ refers to individual matrix element. A non-zero $M_{ij}$ value determines that the $j^{th}$ FG atom contributes to the $i^{th}$ CG particle. 

A few simplifying assumptions can be made immediately. Assume the coordinate transformation is isotropic across all coordinate dimensions. To conserve mass between CG and FG models, each FG particle should be mapped to at least one CG particle. Thus each column in $\textbf{M}$ must have at least one non-zero element. We will restrict our discussion to mapping operators that map each atom to only one CG particle. Therefore, each column in \textbf{M} has exactly one non-zero entry. This restriction is not always imposed for mapping FG to CG particles as seen in some previous works \cite{Fritz2009}. The entries of each row $M_{ij}$ should sum to unity ($\sum_{j=1}^{N_r} M_{ij}=1, \forall i$) because the mass of a CG particle equals the sum of the masses of its constituent atoms. The subset $\mathit{s_i}$ of columns with non-zero entries for the $i^{\text{th}}$ row represent the atoms that are grouped to form the $i^{\text{th}}$ CG particle. For solutions, CG mapping operators can be restricted to the solute and implicit solvent models\cite{Wang2015a}, so that this restriction does not preclude removing solvent. 

The simplifications discussed above have reduced all possible mapping operators from selecting a $N_{\vec{R}}\times N_{\vec{r}}$ matrix to selecting a partition of atoms from which the matrix can be built.
\subsection{Enumeration Schemes for Mapping Operators}
We need to represent molecules as graphs with atoms as nodes and covalent bonds as edges\cite{Balaban1985} in order to continue reducing number of mapping operators. Mapping atoms of a molecule to CG particles while meeting the previously mentioned criteria is equivalent to partitioning the vertex set, \textit{G}, of the molecular graph. Thus, the exhaustive count of mapping operators for a molecule with \textit{n} atoms is obtained by partitioning \textit{G} into disjoint and exhaustive subsets. The total number of ways \textit{G} can be partitioned is calculated by the \textit{n}$^{\text{th}}$ Bell number\cite{Katriel2000}, $\mathit{B_n}$. $B_n$ evaluates the number of ways \textit{n} elements can be partitioned into non-empty subsets. It is given by the expression: $\mathit{B_{n}}=\sum_{k=0}^{n-1} \binom{n-1}{k}\mathit{B_k}$. Since one of the partitions given by $B_n$ represents the FG representation, the exhaustive count for CG mapping operators (referred to as the \emph{count with Bell number}) for a molecule with \textit{n} atoms, is $B_n-1$. Due to its rapid growth rate, the mapping operator count given by Bell numbers becomes intractably large. The mapping operator count can be reduced by joining only bonded atoms. Na\"ively, the number of mapping operators with this limitation is $2^b-1$; $b$ being the total number of bonds. This \emph{na\"ive count} of mapping operators treats all bonds to be unique.

The assumption that all bonds in a molecule are unique is not always true. Molecules have topological symmetries\cite{Chen2004} as a whole, like ethane, or in parts, like the methyl group of methanol \cite{Chen2004}. Bonds in such symmetries are not unique and are equivalent bonds. These equivalent bonds can be defined as edge orbits of the automorphisms of the molecular graph\cite{Balasubramanian1982}. A graph automorphism, which is an isomorphism from a graph to itself, is an operator that permutes vertex labels without changing the graph. When a vertex is re-labeled, it forms an equivalence class with the vertex whose label it adopts. Sets of equivalent vertices are called vertex orbits. The same principle applies to edges, leading to edge orbits. Vertex orbits and edge orbits are our definition of symmetric atoms and bonds respectively. Merging atoms that are joined by equivalent bonds results in duplicate mapping operators. For example, in water the two O-H bonds are equivalent. The mapping operator obtained by removing one O-H bond and merging the corresponding O and H atoms into one bead, will be the same mapping operator that we would get by removing the other O-H bond. Thus the two mapping operators are duplicates. A \emph{mapping operator count without duplicates} due to equivalent bonds can be calculated with the stars and bars theorem of combinatorics \cite{Buhler1994}. The theorem states that  $n$ identical elements can be distributed in $k$ states in $C^{n+k-1}_{k-1}$ ways. In our case, $k$ is 2 since the bonds can be either retained or removed. Thus, the number of CG mapping operators becomes $(\sum_i C^{m_i+k-1}_{k-1})-1$, which is the number of ways $i$ groups of $m_i$ equivalent bonds can be distributed in $k$ states. Note that we have subtracted one from the expression here to exclude the FG representation from the count. The mapping operator count can be further restricted by considering only those that preserve topological symmetries. To preserve symmetry, equivalent bonds are mandated to have uniform states (i.e. all retained or removed) while defining a CG mapping operator. For example, this would disallow removing one C-H bond in a methyl group without removing the other two. The total number of unique \emph{mapping operators preserving symmetry} is $2^{b^{'}}-1$ where $b^{'}$ is the number of unique bonds ($b^{'}\leq b$). It is possible to achieve similar reduction in mapping operator count when a different metric is used to join atoms, like coordination number or spatial proximity instead of the presence of a chemical bond.

\begin{figure}[H]
\includegraphics[width=8.9cm]{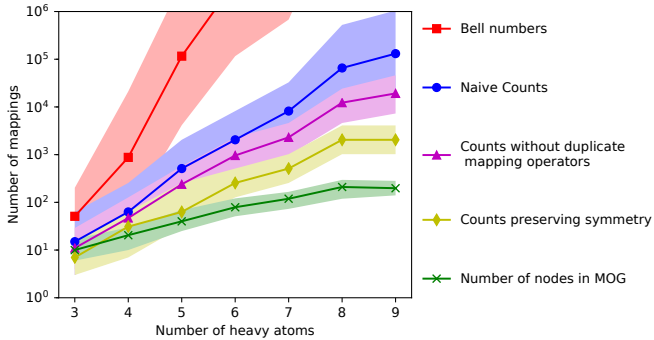}
\centering
\caption{Mapping operator counts given by four counting methods for molecules with heavy atoms in the range 3 to 9. The x-axis denotes the number of heavy atoms in the the molecules. The y-axis is in logarithmic scale. The figure also shows the upper bound of the node count of  the hierarchical graph, called the mapping operator graph (MOG), which encodes multiple mapping operators. The solid lines denote the median values for counts; the upper and lower bounds enclosing the corresponding shaded region denote the upper and lower quartiles respectively. With symmetry and uniform state restrictions, the number of CG mapping operators to be encoded in the hierarchical graph decreases significantly. \label{fig:count-com}}
\end{figure}

We have implemented the four enumeration schemes that have successively fewer mapping operators: \textit{count with Bell number, na\"ive count, count without duplicate mapping operators,\textrm{and} count preserving symmetry}. These have been calculated on 49,889 molecules. For example methanol has $202$ mapping operators using the Bell number. The na\"ive count, assuming all bonds in methanol to be unique, yields a total mapping operator count of $31$. This number includes identical mapping operators and those which do not retain symmetry. Since the methyl hydrogen atoms in methanol are equivalent, the total number of mapping operators without duplicate mapping operators is $15$. The number of mapping operators preserving symmetry is $7$. The semi-log plot in Figure \ref{fig:count-com} shows similar disparity in the order of magnitude of mapping operator counts for other molecules when the four counting methods are used.

\subsection{Hierarchical Graphs Encoding Mapping Operators}
\begin{figure}
\centering
\includegraphics[width=3.33in]{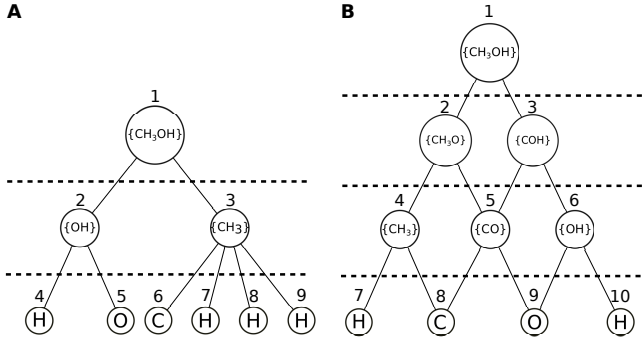}
\caption{(A) Example of an MOT for methanol. (B) Example of an MOG for methanol. Both hierarchical graphs show different levels of granularity, the lowest being the FG representation. The edges show which particles are grouped together in a CG particle. Multiple CG mapping operators are encoded in the MOT and can be extracted with a valid slice. A valid slice is illustrated with the dotted line including nodes 2 and 3. The methanol MOG is a modified version of the methanol MOT allowing multiple parents. The three methyl hydrogen atoms are represented in the MOG with only one hydrogen atom since they are equivalent. The MOG encodes all the seven methanol CG mapping operators that obey symmetry and uniform state restrictions.\label{fig:MOTnMOG}}
\end{figure}

We will introduce two forms of hierarchical graphs: the mapping operator tree (MOT) and the mapping operator graph (MOG). Both encode multiple mapping operators. While the MOG exhaustively encodes all the symmetry preserving mapping operators of a given molecule, MOTs encode a subset of them. MOTs are intended to be applied generally and MOGs are only tractable on small systems. An MOT shows how atoms in a molecule can be grouped to form CG particles at different resolutions. An MOT can be constructed for a molecule so that only symmetry preserving mapping operators are considered and thus serves as the basis for automated selection of mapping operators. Figure \ref{fig:MOTnMOG}(A) shows an illustrative MOT for methanol. We will use the notation \{A$_0$A$_1$A$_2$\} to indicate that atoms A$_0$ though A$_2$ are joined to form a CG particle. Thus single CG bead representation for methanol is denoted by \{CH$_3$OH\}.

An MOT encodes multiple CG mapping operators including mixed-resolution models. For an MOT with $h$ levels, $L^i$ where $i\in [1,h]$, each node in $L^i$ is marked as  $V^i_s$, $s$ being the individual index. The root of the MOT ($V^1_1$) corresponds to the lowest resolution CG representation, mapping one molecule to one CG bead. Subsequent levels of the tree show CG representations in increasing resolution. 
The highest level of the tree ($L^h$) contains the leaf nodes ($V^h_s$) corresponding to atoms in a molecule. Union of $V^i_s$ for each $L^i$ comprises the complete atom set of the molecule. A CG mapping operator can be extracted from the MOT with a valid \textit{tree slice}. A \textit{tree slice} refers to a subset of nodes chosen from the MOT. It can be specified by a binary vector $\vec{x}$ with length equal to the total number of nodes ($V$) in the MOT. If a node is chosen in a slice, the corresponding variable in $\vec{x}$ is set to 1, else it is set to 0. A tree slice is valid only when it includes some representation of all atoms and includes only one node from each of the root-to-leaf paths. A path matrix $\mathcal{P}$ is constructed as a $N_p\times V$ matrix with binary entries, $N_p$ being the number of root-to-leaf paths. An element $p_{ij}$ in $\mathcal{P}$ is $1$ if the corresponding node falls under the path $P_i$. A valid tree slice  $\vec{x}$ obeys $\mathcal{P}\vec{x}=\vec{1}_\mathit{N_p}$, where $\vec{1}_{\mathit{N_p}}$ denotes ones vector of length $N_p$\cite{Xu2013}. Methods like uniform entropy flattening\cite{Xu2013} can be employed to identify a valid and optimal tree slice. 

An MOT cannot encode all mapping operators for most molecules and thus is not unique for a given molecule. 
By choosing valid slices of the methanol MOT in Figure \ref{fig:MOTnMOG}(A), four (\{CH$_3$OH\}, \{CH$_3$\}\{OH\}, \{CH$_3$\}OH, CH$_3$\{OH\}) out of the seven symmetry preserving methanol mapping operators can be extracted. This limitation arises because MOT construction pre-determines the order of bond removal. For instance, in Figure \ref{fig:MOTnMOG}(A) the C-H and the O-H  bonds are removed prior to the C-O bond. Thus the methanol MOT fails to encode these mapping operators: \{CH$_3$O\}H, H$_3$\{COH\}, H$_3$\{CO\}H. The MOT can be adjusted to encode all mapping operators by allowing multiple parents. This results in a mapping operator graph (MOG) that is no longer a tree. Figure \ref{fig:MOTnMOG}(B) illustrates the MOG for methanol.

MOGs can be deterministically and uniquely constructed for a given molecule. The algorithm for MOG construction is provided by Algorithm \ref{alg:algo1}. The source code for MOG construction is available publicly \footnote[0]{https://github.com/ur-whitelab/mapping-graph.git}.
\begin{algorithm}[H]
\caption{MOG Construction:\newline Nodes ($V_g$) and edges ($E_g$) of a molecular graph are taken as input arguments for the MOG function. Using automorphism, the equivalent node classes are identified. The quotient graph of the molecular graph ($V_g , E_g$) is built by partitioning the nodes according to atom equivalent classes. Nodes are added to the MOG only if they are not already present in the MOG. Addition of nodes to all but the first MOG level is followed by addition of directed edges from the node to its children. }\label{alg:algo1}
\begin{spacing}{1.2}
\begin{algorithmic}[1]
\Function {MOG}{$V_g,E_g$}
\State $V_m, E_m \gets \{\}, \{\}$ \Comment{Initalize MOG nodes/edges}
\State $A \gets$ vertex\_orbit$(V_g,E_g)$ \Comment{identify atom equivalent classes}
\State $V_q , E_q \gets (V_g,E_g)/A$\Comment{build a quotient graph of ($V_g,E_g$)}
\For {$i \in V_q$}
\State insert\_node($V_m, E_m, node\leftarrow{}i, level\leftarrow{}$ 0);
\EndFor
\State $d \gets 0$ 
\While {$|\{v \in V_m \,:\, level = d\}| >1$}\Comment{top layer of MOG has more than 1 node}
\For {all node pairs $n_i, n_j \in V_m$ with $level = d$}
\If{$|n_i-n_j|=1 \textbf{ and } |n_j-n_i|=1$} \Comment{check if nodes differ by one element}
\If{($d=0 \textbf{ and } (n_i,n_j) \in E_q) \textbf{ or } d > 0$} 
\State insert\_node$(V_m, E_m, node\leftarrow{}n_i \cup n_j, level \leftarrow{} d +1, children \leftarrow{} n_i, n_j)$
\EndIf

\EndIf
\EndFor
\State $d = d + 1$
\EndWhile
\State \Return $V_m, E_m$
\EndFunction
\end{algorithmic}
\end{spacing}
\end{algorithm}

 Allowing multiple parents introduces multiple paths from the root to leaf nodes and thus the rows in the path matrix must be adjusted by using an OR operation on the paths leading up to the same leaf node to maintain the same equation for valid tree slices. As an illustration, the path matrix ($\mathcal{P}$) for the methanol MOG is shown in Table \ref{tab:a}.
 \begin{table}[H]
\centering
\begin{tabular}{|l|c|c|c|c|c|c|c|c|c|c|}
\hline
Path & V$_1$ & V$_2$ & V$_3$ & V$_4$ & V$_5$ &V$_6$ &V$_7$ & V$_8$ & V$_9$ & V$_{10}$\\
\hline
P$_7$ & 1 & 1 & 0 & 1 & 0 & 0 & 1 & 0 & 0 & 0 \\
\hline
P$_8$ & 1 & 1 & 1 & 1 & 1 & 0 & 0 & 1 & 0 & 0 \\
\hline
P$_9$ & 1 & 1 & 1 & 0 & 1 & 1 & 0 & 0 & 1 & 0 \\
\hline
P$_{10}$ & 1 & 0 & 1 & 0 & 0 & 1 & 0 & 0 & 0 & 1 \\
\hline

\end{tabular}
    \caption{Path matrix for methanol MOG}
    \label{tab:a}
\end{table}
 It must be noted that only one H atom is included from the methyl group as all the three methyl Hs are chemically equivalent. The rule validating a slice still holds. For example, a slice selecting nodes \textit{v$_4$} and \textit{v$_6$} is encoded by the vector $\vec{x}= [0,0,0,1,0,1,0,0,0,0]$. The slice is valid since it satisfies $\mathcal{P}\vec{x}=\vec{1}_\mathit{N_p}$. The slice represents the two particle CG representation of methanol (\{CH$_3$\}\{OH\}). An example of an invalid slice is the selection of nodes \textit{v$_4$}, \textit{v$_5$} and \textit{v$_6$} ($\vec{x}= [0,0,0,1,1,1,0,0,0,0]$) where $\mathcal{P}\vec{x}=[1,2,2,1]$. The slice attempts to capture each of nodes \textit{v$_8$} and \textit{v$_9$} in two CG particles simultaneously. The validation rule not only prevents encapsulating an atom in multiple CG particles but also ensures that a particular atom is chosen to have singular unambiguous resolution by invalidating the selection of a parent and its child simultaneously. The methanol MOG exhaustively encodes the seven mapping operators of methanol which obey symmetry restriction as discussed above. All MOTs for a molecule can be derived from its corresponding unique MOG by removing nodes. Thus the MOG acts as an upper-bound and starting point for MOTs.

\section{Methods}
A fine-grain (FG) simulation of methanol molecules was performed at a density of 0.778 g/cm$^3$ using GROMACS 5.1.3\cite{Mark2015} with the OPLS-AA force field\cite{Jorgensen1996}. The NVT ensemble was maintained at 300K with the Nos\'{e}-Hoover thermostat\cite{Evans1985}. The duration of the simulation was 100 ps. A short range cut-off distance of 12.2 \AA\, was used for Van der Waals interactions, and smooth particle-mesh Ewald\cite{Essmann1995} was used to treat long-range electrostatic interactions. A second  trajectory was produced from the original trajectory by recalculating forces with a topology that excluded all bonded interactions. The recalculated trajectory was used to calculate exclusion forces which correspond to non-bonded interactions. 
R\"{u}hle and Junghans\cite{Ruhle2011} reported exclusion FM performs better than full FM in their study of liquid hexane.

1ns coarse grain (CG) molecular dynamics (MD) simulations were performed corresponding to different CG mappings for methanol using the GROMACS 2016.3\cite{Mark2015} simulation engine and the VOTCA 1.5 package\cite{Ruhle2009}. For each CG mapping, XML files containing the description of the mapping scheme were used to obtain the CG configuration and mapped trajectories  from the FG configuration and the FG trajectory respectively. Force-matching (FM)\cite{Izvekov2005,Noid2007,Ercolessi1994} was employed to obtain tabulated CG potentials from the FG trajectory. FM uses least-squares minimization to calculate CG forces. Both FG and CG simulations were conducted in a box (4.09 nm x 4.09 nm x 4.09 nm) with periodic boundaries. All bonds and angles in the CG simulations were constrained using the SHAKE algorithm\cite{Ryckaert1977}.

To demonstrate the feasibility of applying our hierarchical graph-based method to a larger system, a 200 atom peptide (VPKDGIVAREGSPA) was simulated for 5 ps using the AMBER-99SB-ILDN force field\cite{Lindorff-Larsen2010} and a timestep of 2 fs. The short duration was due to the computational complexity of evaluating the resultant 1250 large atom graphs. The temperature was set at 300 K (NVT). The SPC/E water model\cite{Berendsen1987a} was used to solvate the peptide. Counter ions and NaCl was added to achireve an ionic concentration of 100 mM.

\section{Results}

To demonstrate application of this theory, we have constructed MOTs for methanol and a 14-mer peptide (VPKDGIVAREGSPA) and performed automated selection of CG mapping operators via the uniform entropy flattening method \cite{Xu2013} (UEF). We chose methanol as a test case because it is a commonly used benchmark in coarse-graining literature; the 14-mer peptide shows the scalability of the approach.

The UEF method entails calculating the entropy for each node of the MOT and finding the valid tree slice where the entropy difference between selected nodes is minimized:
\begin{equation} \label{eq:UEF}
\vec{x}_{selected}=\text{arg min} \sum_{V_s,V_t\in MOT}|S(V_s)-S(V_t)|x_sx_t
\end{equation}
In Equation \ref{eq:UEF}, $S$ represents entropy of specific nodes of the MOT, \textit{x} refers to the individual element in the slice vector $\vec{x}$.

\begin{figure}[H]
\centering
\includegraphics[width=8.25cm]{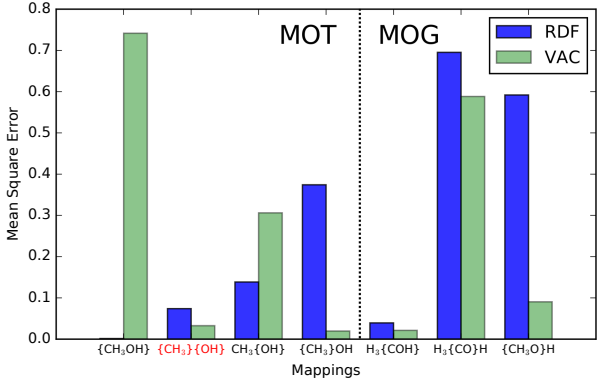}
\caption{Mean square differences of COM-RDFs and velocity autocorrelation functions between FG and seven different methanol CG mapping operators. The mapping operator obtained by applying UEF, highlighted in red, performs best among the four CG mapping operators encoded in the methanol MOT (\{CH$_3$OH\}, \{CH$_3$\}\{OH\}, \{CH$_3$\}OH, CH$_3$\{OH\}). The other mapping operators to the right of the vertical dashed line are those which are encoded in the methanol MOG but not in the methanol MOT. \label{fig:barcomp}}
\end{figure}

\begin{figure}[H]
\centering
    \includegraphics[width=1.1\textwidth]{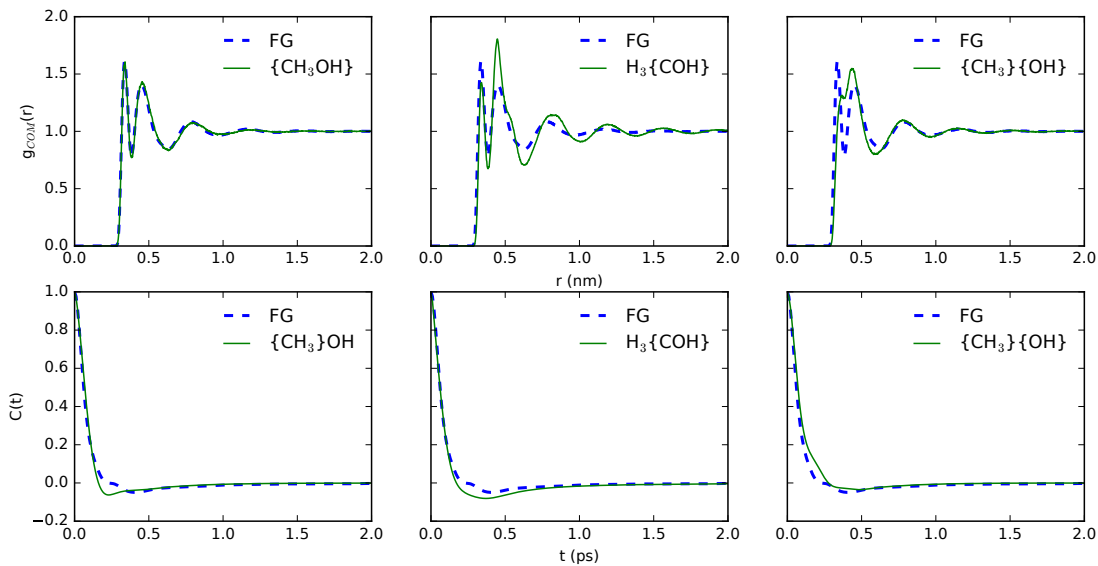}
\caption{Velocity autocorrelation functions (VACFs) and COM-RDFs from the CG MD simulations which best matched those from FG simulation. The COM-RDFs generated with \{CH$_3$OH\}, H$_3$\{COH\} and \{CH$_3$\}\{OH\} methanol CG mapping operators exhibited best fit with the COM-RDF from methanol FG MD simulation. The VACFs generated with \{CH$_3$\}OH, H$_3$\{COH\} and \{CH$_3$\}\{OH\} methanol CG mapping operators exhibited best fit with the VACF from methanol FG MD simulation.\label{fig:subplotComp}} 
\end{figure}

In our methanol example, the negative of the order parameter \cite{Pant2013} was used as an approximation for entropy. All the nodes except the root node were allowed to be selected. The valid slice including all the leaf nodes is a trivial slice corresponding to the FG representation of the molecule; hence it was not considered. The valid slice with nodes 2 and 3 from methanol MOT was selected by the automated method. The chosen slice represents the two bead methanol CG model \{CH$_3$\}\{OH\}. To compare the two bead model with other symmetry preserving mapping operators of methanol, we ran CG simulations of all seven symmetry preserving CG mapping operators of methanol. Although there is no agreed upon method to judge which CG mapping operator is best, here we choose two simple thermodynamic and dynamic functions which have clear definitions in both the CG and FG systems. The first is the center of mass (COM) radial distribution function (RDF)  and the second is the velocity autocorrelation functions (VACF). The mean square differences between the target FG function and the corresponding function from CG simulations provide a quantitative comparison. The mean square differences for different mapping operators are normalized by dividing each element by the norm. Figure \ref{fig:barcomp} compares the normalized mean square differences of CG COM-RDF and VACF of different CG mapping operators for methanol.
The three mapping operators with least mean square differences for RDF and VACF are included in Figure \ref{fig:subplotComp}. It is observed that COM-RDF obtained from one bead (\{CH$_3$OH\}) CG model shows least mean square deviation from the target FG COM-RDF followed by the H$_3$\{COH\} and the \{CH$_3$\}\{OH\} mapping operators. 
VACFs of \{CH$_3$\}OH, H$_3$\{COH\}, \{CH$_3$\}\{OH\} deviate least from the FG VACF. We also see that the two bead model \{CH$_3$\}\{OH\}, chosen by the UEF method performs better than the three other mapping operators encoded in the methanol MOT: \{CH$_3$OH\}, \{CH$_3$\}OH, CH$_3$\{OH\}.

In previous studies comparing CG mapping schemes, the highest resolution CG mapping operator did not necessarily best reproduce the target FG RDF\cite{Markutsya2013,Foley2015,Dallavalle2017}. Foley et. al hypothesized in their study that there might be optimal CG configurations which mirror FG behaviors better than others\cite{Foley2015}. The results obtained underscore the importance of systemically exploring CG mapping operators instead of solely relying on chemical intuition. 

\begin{figure}[H]
\centering
\includegraphics[width=8.25cm]{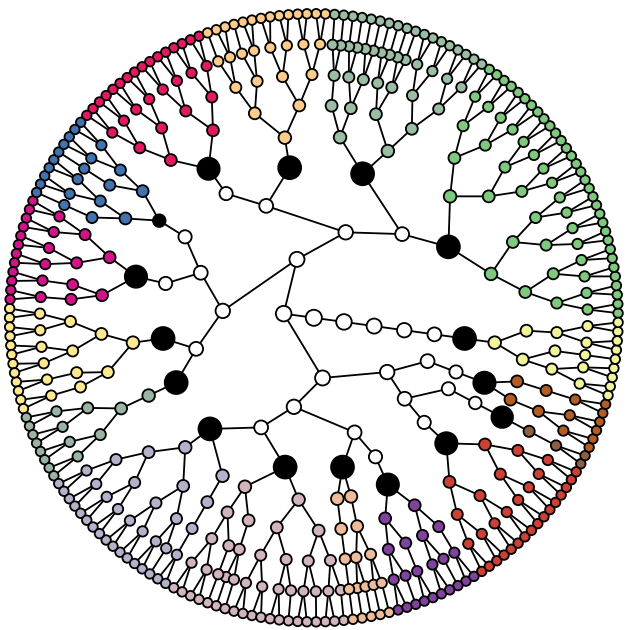}
\caption{An MOT construction for peptide VPKDGIVAREGSPA.  The extraction of a valid slice using UEF method is shown. The chosen nodes are colored black and the children are shown as colored nodes.\label{fig:peptideMOT}}
\end{figure}

The scalability of constructing an MOT and extracting valid mapping operators is demonstrated in Figure \ref{fig:peptideMOT} with the MOT of the peptide VPKDGIVAREGSPA. The MOT construction involved joining particles to one of their four nearest neighbors at every level of the MOT. To  apply  the  flattening,  internal  entropy  of  the  atom  subsets representing each MOT node was  calculated  with  the usual definition of entropy of a probability distribution. The probability distribution is the projected mass onto the principal component of the molecular dynamics trajectory.  Principal component analysis was done over a 5 ps MD simulation trajectory. The principal component number was fixed to explain 95\% of the motion variance.  The  optimal  CG  mapping operator  obtained from UEF is shown  in  Figure  \ref{fig:peptideMOT} where black  nodes  are  the  selection and  their  children  in  the  base  layer  (colored nodes) compose the atom subset.
\section{Conclusion}
We have adapted a hierarchical graph method employed in video segmentation to encode several CG mapping operators. The combinatorial explosion in the number of CG mapping operators is avoided by systematically laying out general restrictions like preserving symmetry or only joining bonded atoms. An algorithm and an illustrative example have been provided to show the construction of a mapping operator graph (MOG) which is capable of exhaustively encoding symmetry preserving mapping operators of a given molecule. Examples were provided to demonstrate how symmetry preserving multi-resolution CG mapping operators can be captured within an MOG. We have also discussed a simpler version of the MOG, called the mapping operator tree (MOT), which can be used to automate the selection of CG mapping operators. Though MOT does not encode all symmetry preserving mapping operators like the MOG, it is still encodes many CG mapping operators and can is more suited for practical use. The theory discussed in this paper provides a basis for automated selection of CG mapping operators by selecting valid tree slices. We have applied the UEF method to the methanol MOT and the peptide MOT as examples of automated slice selection.

\bibliographystyle{ieeetr}
\bibliography{non-bioMolCG-CG-Mapping-Examples,non-bioMolCG,VideoSegmentation-PreviousCG,VideoSegmentation-GraphTheory,VideoSegmentation,Softwares}
\end{document}